\documentclass[twocolumn,british,pra,aps,showpacs]{revtex4-1}
\usepackage[letterpaper]{geometry}
\usepackage{url}
\usepackage{graphicx}
\usepackage{amsmath}
\usepackage{amssymb}
\usepackage{bbold,soul}
\usepackage{esint}
\usepackage{epstopdf}
\usepackage{subfigure}
\usepackage{color}
\usepackage{babel}

\newcommand{\tr}[1]{\textrm{Tr}[ #1 ]}

\definecolor{nblue}{rgb}{0.3,0.3,1.0}
\definecolor{ngreen}{rgb}{0.2,0.7,0.2}
\definecolor{nred}{rgb}{0.9,0.1,0}
\definecolor{norange}{rgb}{0.8,0.5,0}

\begin{document}

\title{Parameter estimation in the presence of the most general Gaussian dissipative reservoir}

\author{Marcin Jarzyna}\email{marcin.jarzyna@fuw.edu.pl}
\author{Marcin Zwierz}\email{marcin.zwierz@fuw.edu.pl}
\affiliation{Faculty of Physics, University of Warsaw, ulica Pasteura 5, PL-02-093 Warszawa, Poland}
\date{\today}

\begin{abstract}\noindent
We analyze the performance of quantum parameter estimation in the presence of the most general Gaussian dissipative reservoir. We derive lower bounds on the precision of phase estimation and a closely related problem of frequency estimation. For both problems we show that it is impossible to achieve the Heisenberg limit asymptotically in the presence of such a reservoir. However, we also find that for any fixed number of probes used in the setup there exists a Gaussian dissipative reservoir, which, in principle, allows for the Heisenberg-limited performance for that number of probes. We discuss a realistic implementation of a frequency estimation scheme in the presence of a Gaussian dissipative reservoir in a cavity system.
\end{abstract}

\maketitle

\section{Introduction}\noindent
The ability to perform precise measurements of physical quantities is of
utmost importance to all branches of science. An example of this is a recent
spectacular detection of gravitational waves by interferometric techniques
\cite{LIGOGW2016, Abbott2016}. The precision of all realistic measurement
setups is limited by various errors caused by detection imperfections and
noise present in the setup. Because of this the state-of-the-art experiments
are performed in such a way as to decrease the impact of all sources of
experimental errors, both the systematic and the statistical ones. This is
done by carefully designing all stages of the experiment and, more
importantly, by actively screening the experimental setup from the
surrounding noise. Strikingly, however, no matter how much effort we put into
our design there is always a physical limitation to any such procedure, as
nature itself sets limits on precision via the principles of quantum
mechanics \cite{Giovannetti2011}.

In most cases, we express the precision limits in terms of the energy, or
equivalently, the average number of probes $\bar{N}$ used in the experiment
as all realistic measurement schemes have to operate on limited resources.
This is because either the power of our source is limited or the experimental
scheme has an upper limit on the energy that it can sustain \cite
{Wolfgramm2012, Taylor2013}. Typically, the higher the number of probes,
hence the energy used, the better is the precision. In the classical regime,
the probes are independent from each other and the precision is bounded by a
standard quantum limit (SQL) or the shot noise scaling $1/\sqrt{\bar{N}}$. On
the other hand, using quantum features of nature, such as entanglement we can
reduce the estimation error and thus improve the precision. In the general
case of parameter estimation, the ultimate bound set by quantum physics is
the so-called Heisenberg limit $1/\bar{N}$ \cite{Giovannetti2006,
Giovannetti2011, Lee2002, Hall2012, Berry2000}. This bound can be achieved
usually only in the decoherence-free case with the use of collective measurements \cite{Seshadreesan2011,
Micadei2015}. Moreover, if we wish to attain the Heisenberg limit, it is
usually necessary to use highly entangled states such as NOON states \cite
{Lee2002, Giovannetti2006, Afek2010, Giovannetti2004}, Holland-Burnett states
\cite{Holland1993}, or twin-beam states \cite{Anisimov2010}.

In real-life experiments, however, we always have to deal with some sort of
decoherence, which can decrease the precision. This issue is  especially
important for the states used to achieve the Heisenberg limit as these states
are often very susceptible to any kind of disturbance. It is known that in
the presence of uncorrelated decoherence, when each probe evolves
independently from the others, the best possible scaling of precision takes
the from of SQL-like scaling $c/\sqrt{\bar{N}}$ \cite{Escher2011,
Demkowicz2012, Knysh2014} (with some notable exceptions \cite{Chaves2013}), where $c$ is a constant, possibly smaller than $1$. Although in some cases we can use error-correction \cite{Dur2014, Kessler2014, Arrad2014} or fast quantum control \cite{Sekatski2016, Sekatski2016a} techniques to restore the Heisenberg-limited scaling  for general decoherence process, the conclusion is that even with entanglement
one can only get a constant improvement over the classical SQL scaling.

In this work, we analyze in detail the problem of phase estimation, employing
a single-mode bosonic probe evolving in the presence of the most general
Gaussian dissipative channel, which, in principle, can be non-Markovian and
noncovariant \footnote{A non-covariant channel is a channel in which the
unitary part of the evolution does not commute with the part describing
decoherence}. This general channel involves important physical channels such
as lossy or additive noise channels as well as more exotic ones, which
involve the evolution of the probe in the presence of a squeezed reservoir.
In addition, we also present the results for a closely related problem of
frequency estimation for the same setup. Towards the end we give an example
of a physical situation in which such an estimation problem may arise.

\section{Estimation theory}\label{sec:basics}
\begin{figure}
\includegraphics[width=\columnwidth]{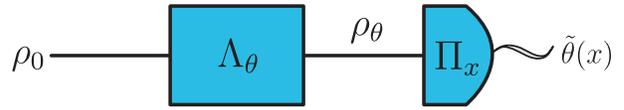}
\caption{(Color online) Standard scheme of parameter estimation. A probe
system prepared in an initial state $\rho_{0}$ is sent through a quantum
channel $\Lambda_{\theta}$, which depends on an unknown value of the
parameter $\theta$ we wish to estimate. The output state $\rho_{\theta}$ is
then subjected to a general POVM measurement $\{\Pi_x\}$. Finally, based on
the measurement outcomes $x$ the value of the parameter is estimated with the
help of an estimator function $\tilde{\theta}(x)$.}
\label{fig:scheme}
\end{figure}

\noindent
The task of inferring any physical quantity from the experiment is described
by quantum estimation theory \cite{Helstrom1976, Holevo1982}. A standard
quantum parameter estimation setup is depicted in Fig.~\ref{fig:scheme}. A
probe system prepared in an initial state $\rho_{0}$ is sent through a
quantum channel $\Lambda_{\theta}$. The evolution under this channel results
in the output state $\rho_{\theta} = \Lambda_{\theta}[\rho_{0}]$, which
depends on an unknown value of the parameter $\theta$ we wish to estimate.
The output state $\rho_{\theta}$ is then subjected to a general quantum
measurement described by a positive operator-valued measure (POVM) $\{\Pi_x\}$. Finally, based on the
measurement outcomes $x$ we can calculate the estimated value of the
parameter with the help of an estimator function $\tilde{\theta}(x)$. The
precision of such an estimation procedure can be quantified with the
root-mean-square error $\Delta\theta = [\langle(\theta -
\tilde{\theta}(x))^{2}\rangle]^{1/2}$, where the averaging $\langle \bullet
\rangle$ is taken with respect to the conditional probability distribution
$p(x|\theta) = \tr{\rho_\theta \Pi_x}$. Note that although the scheme depicted in Fig.~\ref{fig:scheme} is widely used, it is not the most general setup for quantum metrology that one can imagine. A possible extension, created by adding ancillary modes and allowing arbitrary, parameter-independent operations, can also be considered \cite{Dur2014, Kessler2014, Arrad2014, Sekatski2016, Sekatski2016a, Demkowicz2014}; however, here we restrict ourselves to a simple ancilla-free setup.

In order to find the best possible precision one needs to optimize this error
over all stages of the estimation procedure, that is, over all possible input
states, POVM measurements, and estimator functions. This is, in general, a very
hard task. Luckily, the last two optimizations can be avoided by using the
quantum Cram\'{e}r-Rao inequality \cite{Helstrom1976, Braunstein1994}, which
states that for every \emph{unbiased} estimator, the estimation error is
lower bounded by
\begin{equation}\label{eq:CRbound}
\Delta\theta \geq \frac{1}{\sqrt{k F_\theta}}, \quad F_\theta = \tr{\rho_\theta L_\theta^2},
\end{equation}
where $k$ is a number of independent measurement repetitions, $F_\theta$ is
the quantum Fisher information (QFI), and $L_\theta$ is a Hermitian operator
called symmetric logarithmic derivative (SLD), which is defined implicitly via
\begin{equation}
\frac{d\rho_\theta}{d\theta} = \frac{1}{2} (\rho_\theta L_\theta + L_\theta \rho_\theta).
\end{equation}
The quantum Cram\'{e}r-Rao bound given in Eq.~(\ref{eq:CRbound}) is known to
be saturable in the limit of large number of repetitions ($k\to\infty$) with
measurements implementing projections on the eigenvectors of the SLD, whose
results are processed with the maximum likelihood estimator. The remaining
optimization over all possible input states is a much more demanding task as
it necessarily involves the knowledge of the exact form of the quantum
channel $\Lambda_\theta$. In many instances one may refer to matrix product
states optimization \cite{Jarzyna2013} or numerical procedures \cite
{Macieszczak2013, Jarzyna2015}, but still there are some cases that cannot
be efficiently solved with those approaches. Another issue is the
practicality of the optimal states that can be, in principle, derived from the
maximization of the QFI. This is because optimal states are often highly
nontrivial in production and therefore it is also important to consider the
precision offered by experimentally achievable states, such as, for example,
the Gaussian states.

\section{Evolution in the presence of the most general Gaussian dissipative reservoir}\noindent
We begin our study by introducing the basic ingredients of our parameter
estimation procedure. In this paper, we assume that the probe system is
prepared in a single-mode bosonic Gaussian state. A general single-mode
Gaussian state, which represents probe's initial state, can be written as
\begin{equation}\label{eq:StateDens}
\rho_{0} = D(\alpha_{0}) S(r_{0}) \rho_{N_{0}} S^{\dagger}(r_{0}) D^{\dagger}(\alpha_{0}),
\end{equation}
where $\rho_{N_{0}}$ is a single-mode thermal state with average number of
thermal bosons $N_{0}$, $S(r_{0}) = \exp\left[\frac{1}{2}(r_{0}
\hat{a}^{\dagger 2} - r^{*}_{0} \hat{a}^{2})\right]$ is a squeezing operator
and $D(\alpha_{0}) = \exp\left[\alpha_{0} \hat{a}^{\dagger} - \alpha^{*}_{0}
\hat{a} \right]$ is a displacement operator, with $\hat{a}$ and
$\hat{a}^{\dagger}$ being the bosonic annihilation and creation operators.
Hence, the class of Gaussian states includes the thermal states, coherent
states, and, most importantly, squeezed states (more generally, all states that
can be generated with interaction Hamiltonians at most quadratic in $\hat{a}$
and $\hat{a}^{\dagger}$ belong to this class \cite{Braunstein2005}).

A convenient description of Gaussian states is provided by the phase-space
formalism, where any Gaussian state is fully described by only its first and
second moments via the Wigner quasiprobability distribution
\begin{equation}
W_{0}({\bf d}) = \frac{\exp\left[-\frac{1}{2}({\bf d} - \mathbf{\bar{d}}_{0})^{\intercal}\Sigma_{0}^{-1}({\bf d} - \mathbf{\bar{d}}_{0})\right]}{2\pi \sqrt{{\rm det} \Sigma_{0}}},
\end{equation}
where $W_{0}({\bf d})$ is the Wigner function of the initial state $\rho_{0}$
, ${\bf d}=(x, p)$ is a vector of real eigenvalues of the quadrature position
and momentum operators $\hat{x} =
2^{-1/2}\left(\hat{a}+\hat{a}^{\dagger}\right)$ and $\hat{p} = -i
2^{-1/2}\left(\hat{a}-\hat{a}^{\dagger}\right)$, with the first moments
defined via $\mathbf{\bar{d}}_{0} = \langle \mathbf{\hat{d}} \rangle =
(\bar{x}_{0}, \bar{p}_{0})$. The second moments
$\Sigma_0^{ij}=\frac{1}{2}\langle
\mathbf{\hat{d}}_i\mathbf{\hat{d}}_j+\mathbf{\hat{d}}_j\mathbf{\hat{d}}_i\rangle
-\langle\mathbf{\hat{d}}_i\rangle\langle\mathbf{\hat{d}}_j\rangle$ are
arranged into the covariance matrix $\Sigma_0$. For a general single-mode
Gaussian state the covariance matrix, up to phase-space rotations, is given by
\begin{equation}
\Sigma_{0} = \left(N_{0} + \frac{1}{2}\right) \left(\begin{array}{cc}
e^{2 r_{0}} & 0\\
0 & e^{-2 r_{0}}
\end{array}\right).
\end{equation}
Given the above representation, we can easily calculate the average number of
bosons in a single-mode Gaussian state as $\bar{N} =
\frac{1}{2}\left[\left(N_{0} + \frac{1}{2}\right) 2 \cosh 2r_{0} +
|\mathbf{\bar{d}}_{0}|^2 - 1\right]$.

We consider here a situation in which the probe system prepared in a
single-mode Gaussian state $\rho_{0}$ evolves under the action of the most
general Gaussian dissipative channel. Gaussian evolutions belong to a
particularly interesting class of decoherence channels describing many
fundamental processes \cite{Breuer2002, Walls2008}, such as lossy or thermal
evolutions. The dynamics of $\rho_{0}$ evolving under such a channel, in the
interaction picture, can be described by the master equation
\begin{equation}\label{eq:master}
\frac{d \rho_{0}}{d t} = \mathcal{L}(t)\rho_{0},
\end{equation}
where $\mathcal{L}(t)$ is a \textit{Liouville} superoperator given by
\begin{widetext}
\begin{eqnarray}\label{eq:Lsuperoperator}
\mathcal{L}(t) &=& - i \omega H + \frac{\Gamma(t)}{2} \left\{(N + 1) L[a] + N L[a^{\dagger}] + M^{*}D[a]+ M D[a^{\dagger}]\right\},
\end{eqnarray}
\end{widetext}
where $H\rho_{0}=[a^{\dagger}a,\,\rho_{0}]$ describes a free time-independent unitary
evolution of a single bosonic mode $a$ with frequency $\omega$ with the
superoperators $L[o]\rho_{0}=2 o\rho_{0}
o^{\dagger}-o^{\dagger}o\rho_{0}-\rho_{0} o^{\dagger}o$ and
$D[o]\rho_{0}=2o\rho_{0} o-o^2\rho_{0}-\rho_{0} o^2$ accounting for a
coupling to a squeezed thermal reservoir to which
energy is dissipated. Importantly, these superoperators represent also a possible
backaction that the reservoir may have on the state of the system \cite
{Walls2008}. The above master equation takes into account a possible presence
of memory effects that may arise from non-Markovian-type evolutions by
including explicit time dependence in the coupling strength $\Gamma(t)$. We
note that we used the same coupling strength $\Gamma(t)$ for all
superoperators, whereas a fully-general non-Markovian master equation may
have a different coupling parameter for each of the superoperators \cite
{Ferialdi2016}; however, this does not affect the generality of our results.
The parameters $N$ and $M$ in Eq.~(\ref{eq:Lsuperoperator}) are expressed in
terms of the average number
of thermal bosons $N_{\textrm{th}}$, the average number of squeezed bosons
$N_{\rm sq}$, and the squeezing angle $\xi$ of the reservoir \cite{Breuer2002,Walls2008} via
\begin{eqnarray}\label{eq:N}
N &=& N_{\textrm{th}}(2 N_{\rm sq} + 1) + N_{\rm sq}, \\
\label{eq:M}
M &=& (2N_{\textrm{th}}+1) \sqrt{N_{\rm sq}(N_{\rm sq} + 1)} e^{i\xi}.
\end{eqnarray}
These equations are typically combined to provide the relation
\begin{equation}\label{eq:NMrelation}
|M|^2 = N(N+1)-N_{\textrm{th}}(N_{\textrm{th}}+1),
\end{equation}
which states that $N$ and $M$ are not independent quantities. This fact will
become useful in the later analysis.

The formal solution of the master equation in Eq.~(\ref{eq:master}) for time $t$ is given by
\begin{equation}\label{eq:generalsol}
\rho_{\omega} = \mathcal{T} \exp\left[\int_{0}^{t}\mathcal{L}(s) ds \right]\rho_{0} = \Lambda_{\omega} \rho_{0},
\end{equation}
where $\mathcal{T}$ is the time-ordering operator and the superoperator
$\Lambda_{\omega}$ corresponds to the quantum channel depicted in Fig.~\ref
{fig:scheme}, with $\theta = \omega$ making this a frequency estimation
problem. A quick inspection of Eqs.~(\ref{eq:Lsuperoperator}) and (\ref
{eq:generalsol}) shows that we can also write $\rho_{\varphi} =
\Lambda_{\varphi} \rho_{0}$, where $\varphi = \omega t$, meaning that this
theoretical framework allow us to perform phase estimation as well. Finally,
it is important to note that we work here with a tacit assumption of a
perfect reference beam accompanying our single-mode probe \cite{Jarzyna2012}
as estimation employing only a single-mode probe system is not possible.

For a general initial state evolving under a general Gaussian evolution the
final parameter-dependent state usually possesses a very complicated form.
Fortunately, any Gaussian state undergoing Gaussian evolution remains
Gaussian, but its first and second moments are changed \cite
{Braunstein2005,Breuer2002,Walls2008}. The evolution represented by the
$\mathcal{L}$ superoperator results in the first moments
$\mathbf{\bar{d}} = (\bar{x}, \bar{p})$ for the output state $\rho_{\varphi}$
(for the derivation, see Appendix~\ref{app:moments}),
\begin{eqnarray}\label{eq:meanx}
\bar{x} &=& \sqrt{\eta}(\cos \varphi \, \bar{x}_{0} + \sin \varphi \, \bar{p}_{0}), \\
\bar{p} &=& \sqrt{\eta}(-\sin \varphi \, \bar{x}_{0} + \cos \varphi \, \bar{p}_{0}), \label{eq:meanp}
\end{eqnarray}
where $\eta = \exp[-\int_{0}^{t} \Gamma(s) ds]$ is a \textit{time-dependent} dissipation coefficient. The evolved covariance matrix takes a far more complicated form,
\begin{equation}\label{eq:Sigma}
\Sigma =  \eta(\Sigma_{\varphi} - \Sigma_{N}) + \Sigma_{N} + \Sigma_{M},
\end{equation}
where $\Sigma_{N}=(N+\frac{1}{2})\mathbb{1}$, $\Sigma_{\varphi} = R(\varphi)
\Sigma_{0} R^{\intercal}(\varphi)$ is the covariance matrix $\Sigma_{0}$ of
the input state rotated by an angle $\varphi$ \cite{Weedbrook2012} and
$\Sigma_M$ is given in Appendix~\ref{app:moments}. The analogous first and
second moments can also be written for the output state $\rho_{\omega}$ by
replacing $\varphi$ with $\omega t$.

In summary, we wish to stress that the above results fully describe the
evolution of a single-mode Gaussian probe state in the presence of an
arbitrary Gaussian dissipative reservoir including the non-Markovian
reservoirs.

\section{Precision bounds}\noindent
In order to calculate the precision bounds for phase and frequency estimation
we use the expression for the QFI that was derived in
Ref.~\cite{Pinel2013}. In that paper, the QFI for any single-mode Gaussian
state characterized with the first moments $\mathbf{\bar{d}}$ and the
covariance matrix $\Sigma$ is given by
\begin{equation}
\label{eq:QFIGauss}
F_\theta = \frac{1}{2}\frac{\tr{\left(\Sigma^{-1}\Sigma'\right)^2}}{1+\mu^2} + \frac{2 \mu'^2}{1-\mu^4} + \mathbf{\bar{d}}'^{\intercal}\Sigma^{-1}\mathbf{\bar{d}}',
\end{equation}
where $\mu=\frac{1}{2}|\Sigma|^{-1/2}$ is the purity and the primed symbols
simply represent the corresponding first derivatives with respect to the
parameter $\theta$ (in our work this parameter is either $\varphi$ or $\omega$
).

Given the above formula and the definitions of the moments in Eqs.~(\ref
{eq:meanx}), (\ref{eq:meanp}), and (\ref{eq:Sigma}), we obtain expressions for
the QFIs and the corresponding precision bounds for the phase and frequency
estimation in the presence of the most general Gaussian reservoir. Alas, the
resulting precision bounds are very complicated and, of course, crucially
depend on the input state parameters that we have chosen. Therefore, we
present here only the expressions for the asymptotic bounds in the limit of
large average number of probes for two types of input states, that is, for
the optimal Gaussian state, which happens to be the squeezed-vacuum state and
for the coherent state whose performance is usually treated as a benchmark of
a classical behavior. The full expression for the QFI for the general
Gaussian state is given in Appendix~\ref{app:states}. We note also that a
related problem of estimating parameters of a general Bogoliubov
transformation was analyzed in Ref.~\cite{Friis2015}.

\subsection{Phase estimation}
\begin{figure}[t!]
\includegraphics[width=\columnwidth]{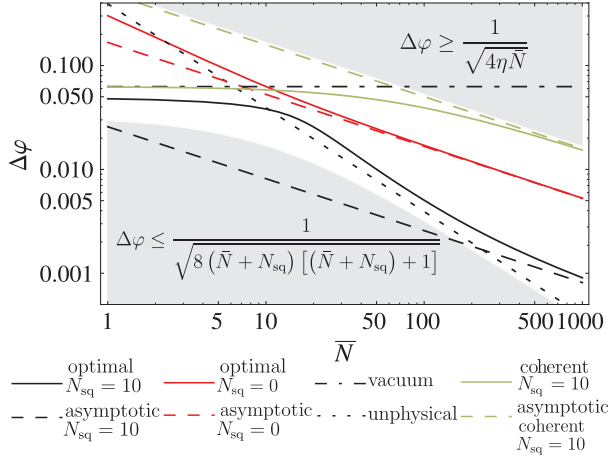}
\caption{(Color online) The exact precision bound for phase estimation in the presence of a purely squeezed dissipative reservoir with $N_{\rm sq} = 10$
and $\xi = 0$ (black solid curve) plotted as a function of the average number
of input probes $\bar{N}$ for $\eta = e^{-\Gamma T} = 0.9$. The green solid
curve is the exact precision bound attainable for an input probe prepared in
the coherent state in the presence of our squeezed reservoir, whereas the red
solid curve represents the exact precision bound for the standard lossy phase
estimation for which $N_{\rm sq} = 0$. The dashed lines of respective colors
are the corresponding asymptotic precision bounds given by Eq.~(\ref
{eq:SqBound}) and Eq.~(\ref{eq:Cohbound}) (with $N_{\rm sq} = 0$ for the red
dashed line). The horizontal black dash-dotted line depicts the precision
obtained for the single-mode vacuum input state for our dissipation model.
The black dotted line represents an unphysical Heisenberg scaling of
$\sqrt{1+\eta^2}/4\eta\bar{N}$, which would hold for an infinitely squeezed
reservoir. The gray shaded areas represent the precision region lying below
the best possible bound of the dissipation-free case (lower one), where for
the purpose of a fair comparison we included into the resource count the
squeezed bosons of the reservoir as if they were a part of the probe system,
and the precision region lying above the sub-SQL-type bound of the standard
lossy phase estimation obtained with coherent input states (upper one).}
\label{fig:sqbounds}
\end{figure}

\noindent
An optimization over all single-mode input Gaussian states, that is, an
optimization over parameters $\bar{x}_{0}$, $\bar{p}_{0}$, $N_{0}$, and $r_{0}$
, reveals that asymptotically, for a large average number of bosons $\bar{N}$
, it is optimal to prepare the probe system in a squeezed-vacuum state.
Importantly, unless environmental squeezing vanishes, our channel is not
covariant; i.e., the phase shift does not commute with decoherence and
therefore unlike most cases analyzed previously in the literature, the QFI
\textit{does} depend on the actual value of $\varphi$. For simplicity, we
assume here that we are interested in estimating the local value of the phase
around $\varphi = 0$. This leads to the asymptotic precision bound for phase
estimation in the presence of the most general Gaussian dissipative
reservoir, which up to the leading order in $N$, is given by
\begin{equation}\label{eq:PhaseBound}
\Delta\varphi \geq \sqrt{\frac{1-\eta}{4\eta\bar{N}}(1 + 2 N - 2|M| \cos\xi)},
\end{equation}
where $\eta = \exp[-\int_{0}^{T} \Gamma(s) ds]$ is the general time-dependent
dissipation coefficient with $T$ denoting the total time of the interaction
of the probe system with the reservoir. The above result states that as long
as $\eta<1$ or $1 + 2N - 2|M|\cos\xi\neq 0$ we obtain at best an SQL-type
scaling with the advantage over the $1/ \sqrt{\bar{N}}$ scaling limited to
the scaling constant. This suggests that if we were able to tune coefficients
$N$ and $M$ in such a way that $1 + 2N - 2|M|\cos\xi= 0$, then the
higher-order terms, proportional to $1/\bar{N}$, would become dominant and we
would obtain Heisenberg-limit-like scaling of
the precision. Unfortunately, we find that such a choice \textit{is not}
possible. To see this, let us assume, without loss of generality, that $\xi=0$
 \footnote{This is the only suitable choice for $\xi$ as it minimizes the
scaling constant of the first sub-SQL-like term in Eq.~(\ref{eq:PhaseBound})
and, incidentally, results in a squeezed dissipative reservoir that has its
squeezing angle aligned with the local value of the phase $\varphi = 0$ to
which our setup is calibrated. The importance of this alignment was also
pointed out in Ref.~\cite{Wu2015}, where phase estimation in the presence of
a squeezed reservoir is studied for the case of the two-level input probe
system.}, which implies the following condition $|M| = N + \frac{1}{2}$.
Combining this condition with Eq.~(\ref{eq:NMrelation}) results in an
equation for $N_{\rm th}$ that has only one (unphysical) solution: $N_{\rm
th} = -1/2$. If we were permitted to set $|M| = N + \frac{1}{2}$ (which, in
fact, is approximately true for an infinitely squeezed dissipative reservoir
characterized with $N_{\rm th} = 0$ and $N_{\rm sq} \rightarrow \infty$),
then the leading order of the asymptotic precision bound would be equal to
$\sqrt{1+\eta^{2}}/4 \eta \bar{N}$, which would represent a far better, Heisenberg-limited
precision than the standard optimal result for lossy phase estimation of $\sqrt{(1 -
\eta)/4 \eta \bar{N}}$ and only a slightly worse than the optimal
scaling for the case without dissipation, that is, the case with $\eta = 1$,
given by $1/\sqrt{8 \bar{N}(\bar{N} + 1)}$ (for details, see Fig.~\ref
{fig:sqbounds}) \cite{Demkowicz2015}.

As the above discussion suggests the most favourable Gaussian dissipative
reservoir that we can consider is a \textit{purely squeezed} dissipative
reservoir with $\xi=0$, for which $N_{\rm th} = 0$ and $N_{\rm sq}$ takes a
fixed value. In such a case, Eq.~(\ref{eq:PhaseBound}) can be rewritten to
the leading order in $\bar{N}$ as
\begin{equation}\label{eq:SqBound}
\Delta\varphi \geq \sqrt{\frac{1-\eta}{4\eta\bar{N}}\left[1 + 2 N_{\rm sq} - 2\sqrt{N_{\rm sq}(N_{\rm sq} + 1)}\right]},
\end{equation}
where we set $N = N_{\rm sq}$ and $|M| = \sqrt{N_{\rm sq}(N_{\rm sq}+1)}$ as
per relation (\ref{eq:NMrelation}). In Fig.~\ref{fig:sqbounds}, we plot the
above asymptotic precision bound for a realistic value of $N_{\rm sq} = 10$
\cite{LIGO2013} (black dashed line) together with the exact precision bound
(black solid curve), that is, the precision bound which is numerically
optimized for each $\bar{N}$ over all single-mode Gaussian input states. In
that figure, we set $\Gamma(s) = \Gamma$. This choice corresponds to the
so-called Markovian dissipation model in which $\eta = e^{-\Gamma T}$
represents the usual loss coefficient. For the purpose of comparison, we
further plot the exact (red solid curve) and the asymptotic (red dashed line)
precision bounds for the standard lossy phase estimation for which $N_{\rm
sq} = 0$. As expected, the presence of squeezed bosons in the dissipative
reservoir, instead of the vacuum, allows for an improved phase estimation for
all values of $\bar{N}$. This improvement is most clearly visible in two
regimes. In the regime of small values of $\bar{N}$, the exact precision
bound is flat and lies below that of the standard lossy phase estimation. In
this regime, the optimal input state is a single-mode coherent state with a
very small admixture of squeezed-vacuum state. In fact, we conclude that for
small $\bar{N}$ one can obtain almost optimal performance even if no input
bosons are sent into the setup at all (see the horizontal black dash-dotted
line in Fig.~\ref{fig:sqbounds}). This behavior is naturally an artifact
specific to our dissipation model: The auxiliary squeezed bosons coming from
the reservoir encode information about parameter and are used for estimation
as well. Furthermore, in the regime of moderate values of $\bar{N}$, which
extends from around $30$ to $140$ bosons, the error in phase estimation
decreases with a Heisenberg-limited fashion. The ratio of the optimal error
of our dissipation model to the optimal error of the perfect dissipation-free
Heisenberg-limited phase estimation is the smallest for $\bar{N} = 66$ and
equal to 1.27.

For the sake of completeness, we finally find that for coherent input states
the asymptotically achievable precision, up to the leading order in $N$, is
bounded from below by
\begin{equation}\label{eq:Cohbound}
\Delta\varphi \geq \sqrt{\frac{1 + 2 (1 - \eta)(N - |M|)}{4 \eta \bar{N}}},
\end{equation}
where for the sake of simplicity we again assumed $\xi = 0$. This formula
indicates that even for coherent input states there exist general Gaussian
dissipative reservoirs for which we obtain an improved performance with
respect to the standard lossy phase estimation. However, unless the
dissipation strength $\eta$ is small, this improvement is minimal (for
details, see Fig.~\ref{fig:sqbounds}). Note also that in the limit $|M|\to N +
1/2$ the bound converges to the noiseless case $\Delta\varphi\geq
1/\sqrt{4\bar{N}}$, indicating that large environmental squeezing can
eliminate the influence of the decoherence.

In summary, we conclude that for an arbitrary Gaussian dissipative evolution
the precision in phase estimation is always bounded from below by an SQL-like
expression. However, for a range of finite values of $\bar{N}$, there \emph
{always} exist a purely squeezed dissipative reservoir with a fixed value of
$N_{\rm sq}$, which allows us to approach a Heisenberg-limited precision very
closely. By increasing the environmental squeezing this range of values
becomes proportionally wider and additionally gets shifted towards larger
$\bar{N}$. Naturally, the above results are also reproduced for fully general
Gaussian dissipative reservoirs with $N_{\rm th} \neq 0$. In the next
section, we determine what kind of improvements can be obtained in the case
of frequency estimation.

\subsection{Frequency estimation}\noindent
As we mentioned in Sec.~\ref{sec:basics}, the lower bound on the error
for frequency estimation is calculated via the quantum Cram\'{e}r-Rao
inequality, Eq.~(\ref{eq:CRbound}). This time, however, the number of
independent measurement repetitions $k$ is not arbitrary. This is because, in
practice, we fix the total time of the experiment $T$ but we can change time
$t$ of each subsequent measurement run. Therefore, the number of repetitions
is equal to $k=T/t$ and since we can modify the interrogation time $t$, the
precision of frequency estimation is given by
\begin{equation}\label{eq:omegaQFI}
\Delta\omega^2 \geq \frac{1}{k F_\omega} = \underset{0 \leq t \leq T}{\rm min} \, \frac{t}{T F_\omega},
\end{equation}
where the quantum Fisher information for frequency $F_\omega$ is calculated
from Eq.~(\ref{eq:QFIGauss}). Therefore, now in order to determine the
fundamental precision of frequency estimation we first optimize the quantum
Cram\'{e}r-Rao inequality over all possible input states and then additionally over
time $t$. Analogously to the case of phase estimation, we are interested in
estimating the local value of frequency around $\omega = 0$, i.e., the detuning from a known frequency.

Since Eq.~(\ref{eq:omegaQFI}) explicitly involves optimization over time, the
optimal precision will crucially depend on the form of the function
$\Gamma(t)$. Here we will focus on a particular class of power functions in
the form of $\Gamma(t)=\Gamma\,t^\beta$, where $\beta$ is a natural number.
This choice is justified since, typically, non-Markovian effects appear on
small time scales and then we may expand almost any, even very complicated,
function $\Gamma(t)$ around $t=0$ and consider only its behavior to the
leading order. For this reason, this power function is one of the most common
choices when considering non-Markovian effects in quantum metrology \cite
{Matsuzaki2011, Chin2012, Smirne2016}. It is important to note here that the units of proportionality constant $\Gamma$ depend on the actual value of $\beta$, but its value depends on the particular design of the experiment and the nature of decoherence process involved.

Under the above assumption an optimization of Eq.~(\ref{eq:omegaQFI}) over
all single-mode input Gaussian states followed by an optimization over
time $t$ leads to an asymptotic precision bound for frequency
estimation in the presence of the most general Gaussian dissipative
reservoir,
\begin{equation}\label{eq:FreqBound}
\Delta\omega^2 T = \frac{1}{4}\left[\left(\frac{1+\beta}{2\beta}\right)^{\beta}\frac{\Gamma (1+2N-2|M|\cos\xi)}{\bar{N}^{2\beta+1}}\right]^{\frac{1}{\beta+1}},
\end{equation}
with the optimal interrogation time for large $N$ given by \footnote{When
optimizing $t/F_{\omega}$, calculated for a squeezed-vacuum state, over $t$
we assume that the optimal time is small in order to simplify the
calculation. Following the optimization we expand $t_{\rm opt}$ around
$\bar{N} \rightarrow \infty$.}
\begin{equation}\label{eq:FreqTime}
t_{\rm opt} =\begin{cases}
\frac{1}{\Gamma \sqrt{\bar{N}(1 + 2 N - 2|M| \cos\xi)}} & \mbox{for} \ \ \beta=0,\\
\left(\frac{1+\beta}{2\beta}\frac{1}{\Gamma(1+2N-2|M|\cos\xi)\bar{N}}\right)^{\frac{1}{\beta+1}} & \mbox{for} \ \ \beta\geq 1.
\end{cases}
\end{equation}
The above bound is asymptotically saturable with squeezed-vacuum states.
Based on the above formulas we can draw similar conclusions as in the case of
phase estimation. In particular, for an unphysical infinitely squeezed
dissipative reservoir characterized with $N_{\rm th} = 0$ and $N_{\rm sq}
\rightarrow \infty$, and with the squeezing angle $\xi$ set to zero, which
all together implies $|M| = N + \frac{1}{2} = N_{\rm sq} + \frac{1}{2}$ the
leading order of the asymptotic precision bound for frequency estimation
would be equal to
$\Delta\omega^2T=\frac{\Gamma^{\frac{1}{\beta+1}}}{16\bar{N}^2}\left(\frac{\beta
}{1+\beta}\right)^{\frac{1}{\beta+1}}(1+e^{\frac{2}{\beta}})$ for $\beta>0$
and $\Delta\omega^2T=\Gamma\frac{1+e^2}{16\bar{N}^2}$ for $\beta=0$ with the
optimal interrogation time $t_{\rm opt}$ scaling to the leading order as
$\left(\frac{1+\beta}{\beta\Gamma}\right)^{\frac{1}{\beta+1}}$ in the former
case and $1/\Gamma$ in the latter \footnotemark[\value{footnote}]. This would
give a Heisenberg-like scaling in estimation error representing a great
improvement over the standard scaling for lossy frequency estimation given by
Eqs.~(\ref{eq:FreqBound}) and (\ref{eq:FreqTime}), with $N = M = 0$. For the
sake of completeness, we note that the optimal scaling for the case without
dissipation, that is, the case with $\Gamma = 0$, is given by Heisenberg-like
scaling expression $\Delta\omega^2 T \geq 1/8 T \bar{N}(\bar{N}+1)$ and is
attained for the optimal interrogation time equal to the total time of the
experiment $t=T$. This reminds us of a similar conclusion that was obtained
for the decoherence-free frequency estimation with the help of Greenberger-Horne-Zeilinger states
\cite{Bollinger1996, Huelga1997}. Let us also note that a similar scaling of
precision as in Eq.~(\ref{eq:FreqBound}) was reported in Ref.~\cite
{Smirne2016}, where the system of $N$ qubits was considered evolving under a
non-Markovian \textit{covariant} channel.

As for the probe system prepared in a coherent state, we find an
asymptotic precision bound for frequency estimation,
\begin{equation}\label{eq:CohFreqBound}
\Delta\omega^2T = \Gamma^{\frac{1}{\beta+1}}\frac{|M|-N}{2\bar{N}}\frac{\left[1+(1+\beta)W(g_\beta)\right]^{\frac{\beta}{\beta+1}}}{(1+\beta)W(g_\beta)},
\end{equation}
with the optimal interrogation time given by
\begin{equation}\label{eq:CohFreqTime}
t_\textrm{opt}=\left[\frac{1}{\Gamma}\left(1 + (1+\beta)W(g_\beta)\right)\right]^{\frac{1}{\beta+1}},
\end{equation}
where $W(\bullet)$ is the Lambert $W$ function \cite{Corless1996}, $g_\beta =
\frac{e^{-\frac{1}{\beta+1}}}{1+\beta}\frac{2(|M| - N)}{2(N - |M|) + 1}$, and
we also assumed that $\xi = 0$. In the simplest case of Markovian evolution
with $\beta=0$, Eqs.~(\ref{eq:CohFreqBound}) and (\ref{eq:CohFreqTime})
simplify to
\begin{equation}
\Delta\omega^2T = \frac{\Gamma}{2\bar{N}}\frac{|M|-N}{W(g_0)},\quad t_{\textrm{opt}}=\frac{1}{\Gamma}(1+W(g_0)).
\end{equation}
This expression represents an improvement over the standard scaling of
frequency estimation with coherent states in the presence of loss:
$\Delta\omega^2T = e \Gamma/4\bar{N}$ with the optimal interrogation time
given by $t_\textrm{opt} = 1/\Gamma$, as per Eqs.~(\ref{eq:CohFreqBound}) and
(\ref{eq:CohFreqTime}) with $N = M \rightarrow 0$ and $\beta=0$.

\begin{figure}[t!]
\includegraphics[width=\columnwidth]{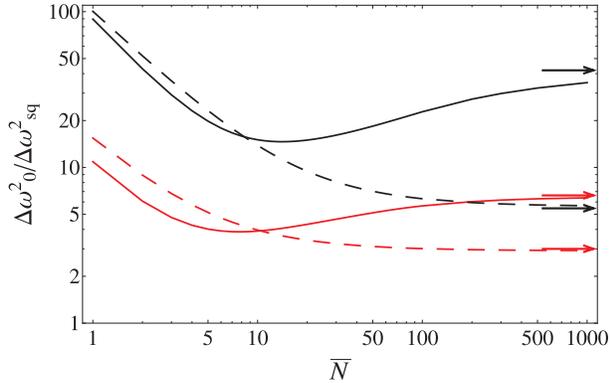}
\caption{(Color online) The ratio of the exact precision bound
$\Delta\omega^2_0$ for the frequency estimation in the case of standard dissipation ($N_{\textrm{sq}}=N_{\textrm{th}}=0$) to the exact precision bound $\Delta\omega^2_{\textrm{sq}}$ obtained in the presence of a
purely squeezed dissipative reservoir ($N_{\textrm{sq}}=10,\,N_{\textrm{th}}=0$) plotted as a function of the average
number of input probes $\bar{N}$ for optimal states with $\beta=0$ (black
solid line) and with $\beta=1$ (red solid line) and for coherent states with
$\beta=0$ (lack dashed line) and with $\beta=1$ (red dashed line). Arrows
of respective colors indicate the asymptotic value of the ratio calculated
from Eqs.~(\ref{eq:FreqBound}) and (\ref{eq:CohFreqBound}).} \label
{fig:frequency_plot}
\end{figure}

In Fig.~\ref{fig:frequency_plot} we plot the gain in the precision of
frequency estimation coming from the squeezing present in the reservoir,
which we quantify by the ratio
$\Delta\omega^2_{0}/\Delta\omega^2_{\textrm{sq}}$ of the precision obtained
for the standard dissipation model for a given $\beta$ to the one obtained with the squeezed reservoir model with the same $\beta$. It can be easily seen that the presence of squeezed bosons in the reservoir is
advantageous in both the Markovian and the non-Markovian cases for both coherent
and optimal states. For optimal Gaussian states the gain is the smallest for
moderate average number of input particles $\bar{N}$ and then increases,
whereas for coherent states it always decreases. The occurrence of a large
gain attainable in the regime of small $\bar{N}$ can be, similarly as in the
case of phase estimation, attributed to the presence of squeezed light coming
from the reservoir. Additionally, in the non-Markovian case the influence of
a squeezed reservoir is weaker when compared to the Markovian one, although
we should keep in mind that the asymptotic precision $\Delta\omega^2T$ in the
former case is better than in the latter. In all cases, however, the gain
saturates at an asymptotic value larger than $1$, indicating that squeezing
in the reservoir is beneficial also in the asymptotic regime of large
$\bar{N}$.

Interestingly, the ratio $\Delta\omega^2_{0}/\Delta\omega^2_{\textrm{sq}}$
does not depend on the actual value of the coupling strength $\Gamma$,
although each precision clearly shows a dependence on this parameter.
Intuitively, this may be understood by referring to Eq.~(\ref{eq:omegaQFI})
and observing that a reparametrization of the QFI $F_\omega=t^2F_{\omega
t}=t^2F_\varphi$ gives us $\Delta\omega^2T=\min_t\frac{1}{t F_{\varphi}}$.
Since $\Gamma$ enters the formula for the QFI of phase estimation only
through $\eta=\exp[-\Gamma t^{\beta+1}/(1+\beta)]$, we see that the time $t$
always has to fulfill $t\sim \Gamma^{-1/(\beta+1)}$ in order for the exponent
to be dimensionless. This, however, means that the dependence on $\Gamma$ of
the precision of frequency estimation is only given through a proportionality
constant $\Delta\omega^2T\sim \Gamma^{1/(\beta+1)}$ irrespective of the
presence of squeezed and/or thermal excitations in the reservoir. Therefore,
the ratio $\Delta\omega^2_{0}/\Delta\omega^2_{\textrm{sq}}$ is the same for
all coupling strengths.

\begin{figure}[t!]
\includegraphics[width=\columnwidth]{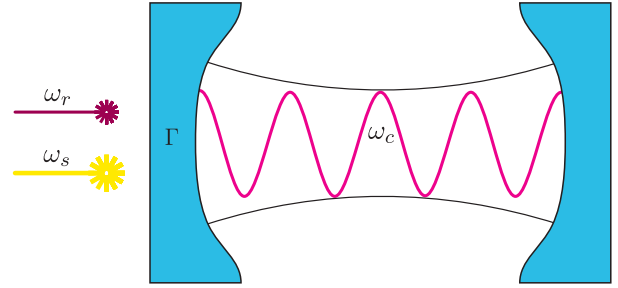}
\caption{(Color online) Schematic of the effective Fabry-P\'{e}rot
interferometer. The cavity mode characterized with the resonant frequency
$\omega_{c}$ is driven by a coherent driving laser with frequency $\omega_{r}$
 and additionally by a broadband squeezed-vacuum light with central frequency
$\omega_{s}$. The effective coupling strength between the coherent and
squeezed fields is given by $\Gamma$.} \label{fig:cavity_optmech}
\end{figure}

\section{Estimating the resonance frequency of a cavity}\noindent
So far our study has been purely theoretical. We now wish to convince the
reader about the practicality of our ideas. To this end, we present a setup
composed of a cavity illuminated by a squeezed beam of light acting as a
dissipative reservoir for a coherent laser field \cite{Gu2013}.

We consider a setup consisting of an effective Fabry-P\'{e}rot cavity, which
is driven by a stabilized coherent laser field with frequency $\omega_r$ and
additionally by a broadband squeezed-vacuum light with central frequency
$\omega_s$ \cite{Gu2013}; the setup is presented in Fig.~\ref
{fig:cavity_optmech}. We wish to sense a detuning $\omega=\omega_c-\omega_r$
of the cavity's resonant frequency $\omega_c$ from the coherent laser field
frequency by inspecting the light field escaping the cavity at frequency
$\omega_r$. The effective coupling strength between the coherent and squeezed
fields is given by $\Gamma$ and we assume that the cavity is \textit
{overcoupled}; i.e., the dissipation inside the cavity is negligible as
compared to the dissipation characterized by $\Gamma$ \cite{Aspelmeyer2014}.
We further assume the evolution to be Markovian.

Since the input signal beam is in a coherent state we can utilize Eq.~(\ref
{eq:CohFreqBound}) to obtain a lower bound on the precision of detuning
estimation. Based on the results presented in the previous section we can
observe the asymptotic gain
$\Delta\omega^2_{\textrm{sq}}/\Delta\omega^2_{0}=\frac{2(|M|-N)}{e W(g_0)}$
in precision resulting from the presence of the environmental squeezing. We
observe similar enhancement in precision for all finite values of the average
number of photons present in the coherent field (see Fig.~\ref
{fig:frequency_plot}).

We note that this model describes a situation, where it is fairly easy to
change the energy of the (coherent) input beam, while the number of photons
in the squeezed beam is fixed. This is very typical for modern
quantum-enhanced optical interferometric experiments in which squeezing
cannot be arbitrarily high; at the moment, the best sources of squeezed light
can produce fields with $N_{\textrm{sq}}\lesssim 10$ \cite{LIGO2011,
LIGO2013, Yonezawa2012}. Finally, let us note that a non-Markovian evolution
with $\beta=1$ is in principle also experimentally feasible in such a model,
as shown in \cite{Jahne2009} \footnote{We note that the non-Markovianity
enters our model not through a time-dependent coupling strength $\Gamma(t)$
but rather through time-dependent noise properties of the squeezed reservoir
\cite{Jahne2009}.}.

\section{Conclusions}\noindent
In conclusion, in the presence of the most general Gaussian dissipative
reservoir, even the squeezed one, the lower bound on the precision of phase
estimation always converges to an SQL-like scaling $c/\sqrt{\bar{N}}$.
However, depending on the type of noise, we can decrease $c$ substantially,
especially in the presence of large squeezing in the reservoir (assuming
thermal excitations are negligible). This enables us to preserve a
Heisenberg-limited precision in the regime of moderate average number of
probes, even though asymptotically Heisenberg scaling requires infinite
environmental squeezing. Importantly, this advantage is present not only for
the optimal Gaussian states, which may be hard to produce experimentally, but
also in the case of coherent states. Similar conclusions may be derived for
frequency estimation in which one has to additionally optimize the protocol
over interrogation time. An interesting open question arising in this context is whether an additional power given to an experimentalist in the form of an error-correction protocol and/or moderate control of the environment would make it possible to preserve Heisenberg-limited scaling also in the regime of finite environmental squeezing. We have also shown that non-Markovian effects,
typically appearing on short timescales, influence the precision of frequency
estimation. Finally, as exemplified by a toy model presented in
the last section, our ideas are not only of theoretical interest but have
interesting practical applications.

\section{Acknowledgments}\noindent
We thank Rafa{\l} Demkowicz-Dobrza{\'n}ski for many insightful discussions
and comments on the manuscript. This work was supported by the European Union Seventh Framework Programme (FP7/2007-2013) projects SIQS (Grant Agreement No. 600645; co-financed by the Polish Ministry of Science and Higher Education) and PhoQuS@UW (Grant Agreement No. 316244)

\appendix

\section{The moments of the output Gaussian probe state}\label{app:moments}\noindent
In order to find the moments for the output state $\rho_{\varphi}$ (or
alternatively $\rho_{\omega}$), we transform the master equation given in Eq.~(
\ref{eq:master}) into a Fokker-Planck-type equation for the Wigner function
$W_{0}(\alpha, \alpha^*, t)$ \cite{Breuer2002,Walls2008,Barnett2003}:
\begin{widetext}
\begin{eqnarray}
\frac{\partial W_{0}(\alpha, \alpha^*, t)}{\partial t} &=& \left(\frac{\Gamma(t)}{2} + i \omega \right) \frac{\partial}{\partial \alpha}[\alpha W_{0}(\alpha, \alpha^*, t)] + \left(\frac{\Gamma(t)}{2} - i \omega \right) \frac{\partial}{\partial \alpha^{*}}[\alpha^{*} W_{0}(\alpha, \alpha^*, t)] \nonumber \\
&+& \frac{\Gamma(t)}{2} \left[ M\frac{\partial^2}{\partial\alpha^2} + M^{*}\frac{\partial^2}{\partial\alpha^{*2}} + 2 \left(N + \frac{1}{2}\right)\frac{\partial^{2}}{\partial \alpha \partial \alpha^{*}}\right]W_{0}(\alpha, \alpha^*, t).
\end{eqnarray}
\end{widetext}
We used here the representation of the Wigner function in terms of the
complex amplitudes $\alpha = 2^{-1/2} (x + ip)$ and $\alpha^{*} = 2^{-1/2} (x
- ip)$ because it allows us to write the following set of \textit{uncoupled}
ordinary differential equations \cite{Barnett2003}:
\begin{widetext}
\begin{eqnarray}
\frac{d}{dt} \langle \alpha^{n}(t) \alpha^{*m}(t) \rangle &=& - n \left(\frac{\Gamma(t)}{2} + i \omega \right) \langle \alpha^{n}(t) \alpha^{*m}(t) \rangle - m \left(\frac{\Gamma(t)}{2} - i \omega \right) \langle \alpha^{n}(t) \alpha^{*m}(t) \rangle \nonumber \\
&+& \frac{\Gamma(t) M}{2} n(n-1) \langle \alpha^{n-2}(t) \alpha^{*m}(t) \rangle + \frac{\Gamma(t) M^{*}}{2} m(m-1) \langle \alpha^{n}(t) \alpha^{*m-2}(t) \rangle \nonumber \\
&+& \Gamma(t) \left(N + \frac{1}{2}\right) n m \langle \alpha^{n-1}(t) \alpha^{*m-1}(t) \rangle.
\end{eqnarray}
\end{widetext}
Here the averages are calculated with respect to $W_{0}(\alpha, \alpha^*, t)$
. Based on the above system of equations, we are able to calculate the
required moments of $\alpha(t)$ and $\alpha^{*}(t)$. The formal solutions are
given by
\begin{eqnarray}
\langle \alpha(t) \rangle &=& \sqrt{\eta} \langle \alpha_{0} \rangle e^{- i \varphi}, \\
\langle \alpha^{*}(t) \rangle &=& \sqrt{\eta} \langle \alpha^{*}_{0} \rangle e^{i \varphi}, \\
\langle \alpha^{2}(t) \rangle &=& \eta \langle \alpha^{2}_{0} \rangle e^{- 2i \varphi} + \eta M \mathcal{C} e^{- 2i \varphi}, \\
\langle \alpha^{*2}(t) \rangle &=& \eta \langle \alpha^{*2}_{0} \rangle e^{2i \varphi} + \eta M^{*} \mathcal{C}^{*} e^{2i \varphi}, \\
\langle \alpha(t) \alpha^{*}(t) \rangle &=& \eta \langle \alpha(0) \alpha^{*}(0) \rangle + (1 - \eta) \left(N + \frac{1}{2}\right), \nonumber \\
\end{eqnarray}
where
\begin{equation}
\mathcal{C} = \int_{0}^{t} \exp\left[\int_{0}^{s} \Gamma(\tau) d\tau \right] \Gamma(s) e^{2 i \omega s} ds
\end{equation}
and $\eta = \exp[-\int_{0}^{t} \Gamma(s) ds]$. We can use these solutions to
calculate the moments of the quadrature position and momentum operators for
the output state $\rho_{\varphi}$ (or alternatively $\rho_{\omega}$). The
first and the second moments (which are arranged into a covariance matrix
$\Sigma$) are given in the main text in Eqs.~(\ref{eq:meanx}), (\ref
{eq:meanp}), and (\ref{eq:Sigma}). Here we only present a rather complicated
form of the $\Sigma_M$ matrix, which enters the definition of the covariance
matrix $\Sigma$,
\begin{equation}
\Sigma_M=\eta|M||\mathcal{C}|\left(\begin{array}{cc}
\cos(\theta-2\varphi) & \sin(\theta-2\varphi)\\
\sin(\theta-2\varphi) & - \cos(\theta-2\varphi)
\end{array}\right),
\end{equation}
where we have defined $\theta$ as a phase of $M\mathcal{C}$; i.e., $M\mathcal{C}=|M||\mathcal{C}|e^{i\theta}$.

\section{The exact precision for pure Gaussian states}\label{app:states}\noindent
The exact formula for the QFI of phase estimation with a pure Gaussian probe
state evolving under an arbitrary Gaussian evolution characterized by the
overall loss coefficient $\eta$ and parameters $N$ and $M$ is given by
\begin{widetext}
\begin{eqnarray}\label{eq:app_QFI_exact}
F_{\varphi}&=&\frac{4\eta^2(\bar{n}-2M)^2}{\left\{1+2(1-\eta)(N+M)+2\eta[\bar{n}+\sqrt{\bar{n}(\bar{n}+1}]\right\}\left\{1+\frac{1}{1+4(1-\eta)(N+M)[(1-\eta)(N-M)+\eta\bar{n}]+4\eta(1-\eta)[N+\bar{n}(N-1)]}\right\}}+\nonumber \\
&+&\frac{4\eta(\bar{N}-\bar{n})}{1+2(1-\eta)(N-M)-2\eta[\sqrt{\bar{n}(\bar{n}+1)}-\bar{n}]},
\end{eqnarray}
\end{widetext}
where $\bar{N}$ is the total average number of photons and
$\bar{n}=\sinh^2r_0$ is the number of squeezed photons in the input state and
we have assumed that $\xi=\varphi=0$ for simplicity. The first part of the
above expression is due to the squeezing present in the input, whereas the
second one quantifies the influence of the initial displacement. From Eq.~(
\ref{eq:app_QFI_exact}) one can easily obtain the QFI for coherent state
input as well as expressions for frequency estimation.
\end{document}